# FEASIBILITY OF 5G MM-WAVE COMMUNICATION FOR CONNECTED AUTONOMOUS VEHICLES

**Sakib Mahmud Khan***
**Ph.D. Candidate**
Clemson University
Glenn Department of Civil Engineering
351 Fluor Daniel Engineering Innovation Building, Clemson, SC 29634
Tel: (864) 569-1082, Fax: (864) 656-2670
Email: sakibk@g.clemson.edu

**Mashrur Chowdhury, Ph.D., P.E., F.ASCE**
**Eugene Douglas Mays Endowed Professor of Transportation and Professor of Automotive Engineering**
Clemson University
Glenn Department of Civil Engineering
216 Lowry Hall, Clemson, South Carolina 29634
Tel: (864) 656-3313   Fax: (864) 656-2670
Email: mac@clemson.edu

**Mizanur Rahman**
**Ph.D.**
Glenn Department of Civil Engineering, Clemson University
351 Flour Daniel, Clemson, SC 29634
Tel: (864) 650-2926; Fax: (864) 656-2670;
Email: mdr@clemson.edu

**Mhafuzul Islam**
**Ph.D. Student**
Glenn Department of Civil Engineering, Clemson University
351 Flour Daniel, Clemson, SC 29634
Tel: (864) 986-5446; Fax: (864) 656-2670;
Email: mdmhafi@clemson.edu

*Corresponding author

**ABSTRACT**

The internet-of-things (IoT) environment holds different intelligent components networked together and will enable seamless data communication between the connected components. Connected autonomous vehicles or CAVs are major components of the IoT, and the smooth, reliable, and safe operation of CAVs demands a reliable wireless communication system, which can ensure high connectivity, high throughput and low communication latency. The 5G millimeter-wave or mm-wave communication network offers such benefits, which can be the enabler of CAVs, especially for dense congested areas. In this research, we have evaluated the 5G mm-wave and Dedicated Short Range Communication (DSRC) for different CAV applications in a Network Simulator-3 (ns-3). For the CAV applications, we have evaluated the end-to-end latency, packet loss, and data rate (for both the CAV receiver and transmitter) of the 5G mm-wave. We have found that 5G mm-wave can support CAV safety applications by ensuring lower latency compared to the required minimum latency of 200 milliseconds for the forward collision warning application. For mobility applications, we have found that 5G mm-wave can support multiple CAVs with a high data receive rate, which is enough for real-time high definition video streaming for in-vehicle infotainment, with mean packet delay of 13 milliseconds. The findings from this study establish that 5G mm-wave can be the enabler of future CAVs in congested areas. Using the evaluation framework developed in this paper, public agencies can evaluate 5G mm-wave to support CAVs in congested areas, such as heavy pedestrian areas like downtown, commercial zones, under their jurisdiction.

## INTRODUCTION

Innovations in the domain of wireless communication technology have supported the massive growth of interconnected personal devices and intertwined sensory equipment owned by different agencies. However, the everlasting emergence of connected devices has increased the demand on the research community to develop a wireless communication enabler to support continuous and reliable data flow between the connected devices. Many are thinking that the fifth-generation wireless communication network or 5G will be the enabler to connect and support numerous connected devices under the broad umbrella of the internet-of-things or IoT (1). IoT refers to the networked integration of highly distributed day-to-day devices via embedded systems (2). 5G wireless communication once deployed has the potential to provide better communication reliability to the IoT interconnected devices compared to the existing fourth-generation wireless communication (4G). In transportation, connected autonomous vehicles (CAVs) are one of the major components of IoT, and numerous safety, mobility and environmental benefits can be derived from CAV applications. In this paper, we have studied how CAVs can leverage the 5G wireless communication. As identified in previous literature, the advantages of 5G over 4G are: increased spectrum allocation, availability of directional beamforming antennas used at both 5G enabled base stations and mobile entities, increased capacity to aggregate numerous simultaneous users within the coverage area, highly increased bit rates within increased proportions of the 5G coverage areas, and lower cost of infrastructure (3).

For 5G to increase communication network reliability, throughput, and connectivity in the IoT environment, researchers are showing more interest than before on spectrums higher than 6 GHz. The under 6 GHz spectrum is already divided into different Long Term Evolution (LTE) spectrums. The unused spectrums above 6 GHz are the potential areas that can provide higher throughput for the increasing number of CAVs in the future, especially in dense urban areas. The term 'mm-wave' in '5G mm-wave' simply refers to the wavelengths within the range of one to ten millimeters (4). Compared to recent 4G, the carrier frequency of 5G mm-wave allows for the broadcast of highly increased data rates while reducing the communication latency. Due to this inherent capacity offered by 5G mm-wave for both backhaul links (within multiple base stations) and access links (within the base station and end users), it can support the CAV environment where multiple data-intensive safety, mobility, and in-vehicle infotainment applications can run in numerous CAVs in highly congested areas. The CAVs can be equipped with in-vehicle devices with dynamic beamforming-enabled multi-element antennas to connect with the mm-wave base stations. The coverage area of the mm-wave network is limited compared to LTE. To overcome this limited mm-wave coverage area issue, the base stations should be closely spaced. Wireless network operators are continuously reducing the cell coverage areas to enhance relaying, implement cooperative multiple-input-multiple-output with multiple antennas at the receiver and sender, and reduce inference (3). As a result, the number of base stations will increase in the future mm-wave based deployments, which will be applicable to the future CAV environment where an unprecedented number of CAVs will demand multi-gigabit/second to support CAV applications and to stream real-time in-vehicle infotainment components. The applications of mm-wave communication have been studied earlier. For the mobile communication system, Dehos et al. (2014) have identified mm-wave as the primary technology for next-generation communication (5). Mastrosimone and Panno (2015) have studied the performance of hybrid mm-wave and LTE access links and compared the hybrid system against pure LTE-based access links, and they have found an increased throughput of 33% using the hybrid mm-wave and LTE access links compared to the only LTE scenario (6).

Recent pilot deployment sites in the US to evaluate connected transportation applications are mostly equipped with Dedicated Short-range Communication (DSRC)-based devices, and in a few cases, a large area is wirelessly connected with combined DSRC and 4G LTE to support the safety and mobility applications. However, cellular network service providers are pursuing the plan to deploy a 5G network commercially. A comparative investigation between 5G mm-wave and DSRC for different CAV applications is missing. The International Telecommunication Union Radiocommunication Sector has planned to finalize all specifications of 5G by 2020 (7), based on pilot real-world tests conducted in various countries. Before conducting the real-world tests, studies conducted in a simulation environment can provide extensive understandings of 5G-mm-wave benefits for different CAV applications in multiple scenarios, such as congested/uncongested and low/high penetration of CAVs. Several studies have investigated the 5G mm-wave network for vehicular communication. However, the feasibility of the 5G network has not been thoroughly studied for the environment that considers CAV application requirements. The research objective of this paper is to investigate the potential benefits of a 5G mm-wave network for CAV applications. Using the evaluation framework developed in this paper, public agencies can evaluate the use of 5G mm-wave to support CAVs in congested areas.

## WHY 5G MM-WAVE CONSIDERED FOR CAV?

CAVs have different in-vehicle sensors, which include cameras, lidar, GPS, and ultrasonic sensors for safe and reliable operation. A massive amount of data is generated from these sensors. For example, Huang et al. (2010) have discussed the dataset generated by the MIT autonomous vehicle during the 2007 DARPA urban challenge (8). The autonomous vehicle had lidar, cameras, and a GPS sensor installed. According to the dataset, the generated data is almost 4 GB/mile (8). In the connected environment, these massive sensor generated data can be crowdsourced in real-time. Acquiring 3D high definition maps is critical for safe CAV operations. Another data-intensive CAV application is the in-vehicle infotainment system. Streaming high-quality video to a massive number of CAV users in a congested area demands a high data rate for wireless communication. While running all these data-intensive CAV applications, the communication network should also ensure minimum latency for safety-critical applications. The 5G mm-wave thus becomes a viable communication option for CAVs. In the US, wireless communication providers like AT&T, Verizon, and T-Mobile have included 5G mm-wave in their overall 5G communication deployment plan, which would potentially lead to the scenario of accessing the 5G mm-wave network for CAV applications in future. Figure 1 shows the 5G mm-wave enabled CAV scenario for an urban congested area where a data center is connected with the mm-wave base station either: (a) directly with the fiber optic network or (b) via macro base stations. The macro base station can be wirelessly connected with the data center with microwave connection for non-line-of-sight scenarios.

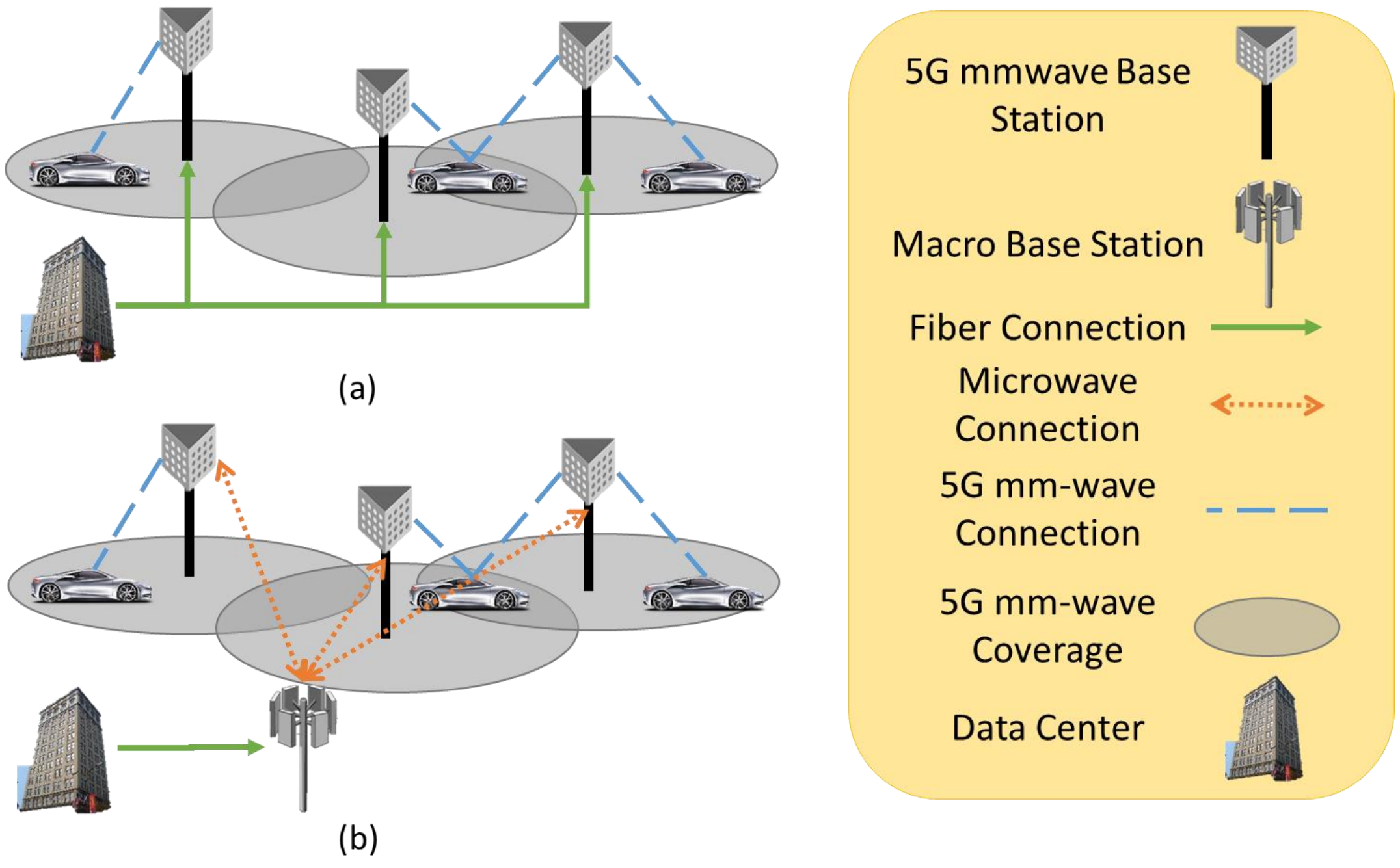

**FIGURE 1 5G mm-wave network in urban environment**

## LITERATURE REVIEW

### 5G status in US and worldwide

In order to provide a higher data rate, the telecommunications industry and academia have been involved in research with the goal of improving the spectral communication efficiency by deploying more base stations of 4G LTE. Even with the significant enhancement in heterogeneous networks (HetNet) that include 4G LTE, DSRC, and wifi, these technologies are falling behind to meet user data requirements. The advent of the 5G era, which can support thousands of times greater data traffic demands, is expected to solve the increasing demand of data traffic. To date, 4G cellular communication systems have been adopted in the US and worldwide. Every 10 years, a new generation of emerging communication technologies is replacing the old technologies since 1980: first generation analog FM cellular systems in 1981, second generation digital technology in 1992, 3G in 2001, and 4G LTE-A in 2011 (9).

Currently, targeting 2020, the race to search for innovative solutions to enable the 5G era has been ongoing. In early 2013, the European Commission announced the investment of €50 million for 5G research in multiple projects, such as METIS. The Chinese government led IMT-2020 Promotion Group in February 2013, and the Korean government led the 5G Forum in May 2013. The vision of 5G is to offer a gigabits-per-second experience to end users. The standardization of 5G specifications by the Third Generation Partnership Project (3GPP) and the formal ratification of 5G standards by the International Telecommunication Union are still ongoing to make the vision of the high data rate of 5G a reality (3, 10).

In the US, the telecommunication industry is being pushed to deploy 5G as a fully operational network (11). However, telecommunication industries such as ATT&T, Verizon, and Sprint are still operating 5G in a testing phase. AT&T is leading by deploying 5G networks in 23 metro areas including Atlanta, Boston, New York, Chicago, San Francisco and Houston. Verizon plans to deploy commercial 5G network in California in 2018 (12).

### 5G Application in vehicular networks

The following issues pose challenges for the reliability of wireless communication in a CAV system for safety, mobility and environmental applications: (i) the dynamic network topology of vehicular communication changes very frequently due to high vehicle mobility and results in frequent data flow disconnections (13), (ii) cross-channel interference in vehicle-to-vehicle communication introduces packet drop rates when two adjacent channels are operated simultaneously (14). Specifically in a higher vehicle density scenario, the intensity of channel contention among vehicles increases significantly, which results in a higher transmission collision rate and a larger channel access delay (15).

To overcome these challenges, 5G mm-wave is expected to expand and support various application scenarios, such as enhanced Mobile Broadband (eMBB), Ultra Reliable Low Latency Communication (URLLC), and massive Machine Type Communication. eMBB is designed for the high data rate mobile broadband services which require seamless data access both indoors and outdoors. In addition, URLLC is designed for applications that require stringent latency and reliability requirements in highly mobile vehicular communications to enable the CAV network. Furthermore, 5G should be backward compatible with current LTE, and its control plane should be the same as LTE. The “5G/IMT-2020 Standing Committee” reports the standards and projects of plausible relevance to fifth generation wireless communication (5G) technologies (16). The IEEE 802.11 Working Group (WG) develops standards for IEEE 802.11ad for 5G standards with the follow-up project P802.11ay, which supports extremely high individual throughput in the mm-wave bands as part of 5G technologies. Studies conducted by Choi et al. (2016) and Polse et al. (2017) have shown that 5G mm-wave can be applicable for mobile nodes (e.g., connected vehicles) because of its high communication bandwidth with a gigabit/sec data rate and low latency communication delay (9, 17).

## RESEARCH METHOD

### Study Area Selection

To evaluate the efficacy of 5G mm-wave, we have selected one heavily congested road from Greenville County, South Carolina, which is Woodruff Road. Over the timespan of 30 years (1960-1990), the surrounding area has been transformed into a commercial hub from its early rural beginnings with the current average daily traffic being 40,000 cars. According to a survey conducted in 2017, among 4198 respondents, 67% and 81% identified the operational condition as ‘very congested’ for weekdays and weekends, respectively (18). TABLE 1 shows the issues of the respondents with the traffic condition on Woodruff Road, which would justify the implementation of CAV safety and mobility applications in this corridor.

**TABLE 1 Existing Traffic Issues with Woodruff Road (18)**

| Current Issues with Woodruff Road | Percentage of over-concerned respondents |
|---|---|
| Safety | 64% of 4198 respondents |
| Traffic operations with existing traffic signal | 63% of 4198 respondents |

Using the Simulation of Urban Mobility (SUMO) simulator, we have extracted the tracefile from SUMO to be used in Network Simulator-3 (ns-3) to investigate the performance of wireless communication. The tracefile has information about each vehicle node, position, and speed, as well as the timestep when the vehicle data is captured.

### CAV Application Selection

To evaluate the performance of 5G mm-wave, two CAV applications are selected, which are: 1) forward collision warning and 2) traffic data collection. In the forward collision warning scenario, once any CAV faces collision it needs to immediately broadcast the ‘Forward Collision Ahead’ warning to the following CAV. For the CAV safety applications, the time between the collision warning from one vehicle to the next needs to be less than 200 milliseconds (ms) (19). As shown in FIGURE 2, the leader CAV, when it suddenly collides with any other vehicles or faces any in-vehicle malfunction, sends the collision warning to the mm-wave base stations, and all the follower CAVs receive the data from the base station. In the DSRC environment, the follower vehicle receives the warning immediately from the leader CAV, as both are equipped with DSRC. Findings from this study will support whether 5G mm-wave can support CAV safety applications meeting the less than 200 ms latency requirement. From the analysis of the second CAV application called “CAV data collection”, the feasibility of CAV data transmission and reception data rate using DSRC and 5G mm-wave will be evaluated.

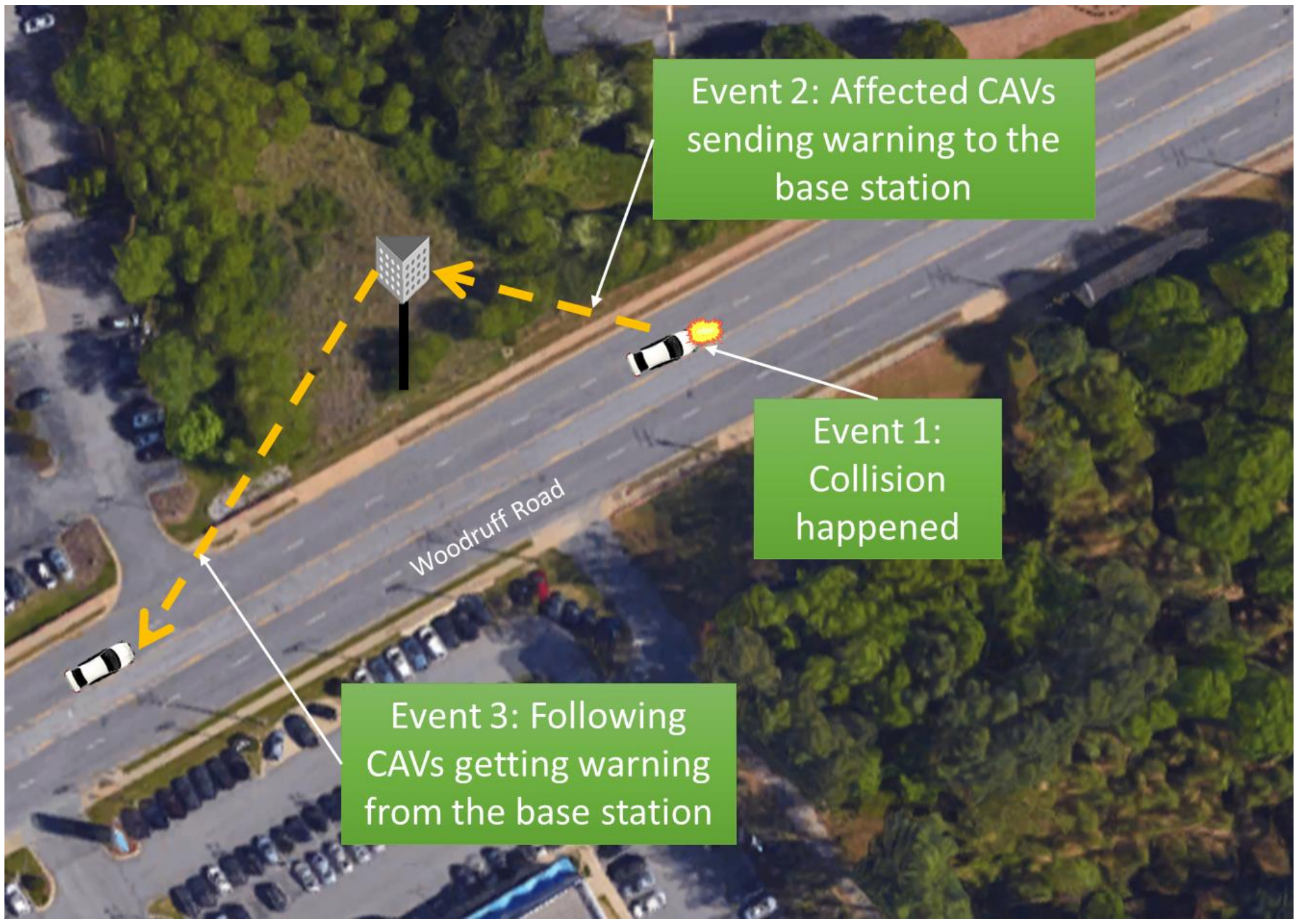

**FIGURE 2 Forward collision warning in 5G mm-wave environment**

## Communication Model Selection

*Wireless Communication Option 1: DSRC*

To investigate the DSRC communication system's performance for both the safety and data collection applications, we have selected the ns-3 simulator. Ns-3 is already established as a reliable communication network simulator, and it provides different wireless communication testing modules including LTE, wifi, DSRC etc. The wireless access in vehicular environment (WAVE) module in ns-3, developed following the IEEE 802.11p, IEEE 1609 and SAE J2735 standards, is used to simulate the DSRC environment. For the DSRC environment, a 5.9 GHz frequency spectrum is used. Using this spectrum, CAVs sent basic safety message data of about 200 bytes every one-tenth of a second (i.e., 10Hz frequency). In the DSRC environment, data generated by the CAVs will be captured by the roadside units (RSUs), which are stationary units installed on the side of roads. While simulating the CAVs and RSUs in ns-3, we have first installed mobility models in these nodes. The mobility of CAV nodes are assigned based on a SUMO trace file, and the RSU nodes are treated as stationary nodes. After installing the mobility model, we have installed the Ad hoc On-Demand Distance Vector (AODV) routing protocol, and we have assigned IP addresses to each node so that we can identify the communicating nodes later. Later, we have defined the physical layer and data link or MAC layer, and we have installed the channel propagation model and the propagation loss model. At last, the UDP application layer protocol is installed. As shown in TABLE 2, we have used the Friis propagation loss model and Nakagami fast fading model while simulating DSRC.

**TABLE 2 Consideration for Wireless Communication for CAVs**

| **Wireless Communication Modules** | **Considerations** |
|---|---|
| **DSRC** | Frequency Band: 5.9 GHz<br>Packet size: 200 bytes (for Basic Safety Message)<br>Data rate: 160 Kbps<br>Propagation Loss Model: Friis<br>Fast Fading Model: Nakagami<br>Routing Protocol: Ad hoc On-Demand Distance Vector (AODV) routing<br>Transmission and Receiving Gain: 1 dB (20) |
| **5G mm-wave** | Frequency Band: 28 GHz<br>Packet size: 1400 bytes (including Basic Safety Message)<br>Data rate: 4000 Kbps<br>Blockage Model: Enabled<br>Fading Model: Small-scale fading<br>PHY layer: Hybrid Automatic Repeat Request enabled (To provide transmission robustness)<br>MAC Layer: Round Robin Scheduler |

*Wireless Communication Option 2: 5G mm-wave*

In order to simulate the 5G mm-wave module in ns-3, we have used the mm-wave module developed by New York University (21). This recently introduced mm-wave module offers a wide range of frequency-based (i.e., 6-100 GHz) testing. In this paper to simulate the 5G mm-wave, we have used hybrid automatic repeat request based retransmission, which helps to do fast retransmission of data packets and increases the probability of successful decoding at the receiving end. We have also considered signal blockage because mm-wave signals are highly susceptible to blockages due to external objects like human bodies or vehicles. TABLE 2 contains the list of all components considered for this 5G module. Similar to the DSRC testing, all mobile nodes (i.e., CAVs) are created based on the SUMO generated tracefile. Later, static nodes (i.e., base stations) are created in ns-3. After adding the mobility, routing, mm-wave and IP layer stacks, we have tested the UDP-based application to evaluate the communication system for CAV safety and data collection applications.

## Performance Measurements of Communication Options

For the CAV forward collision application, we have measured the end-to-end delay for both DSRC and the 5G mm-wave. For the UDP-based safety application, the delay is measured from the time the leader CAV sends the data after experiencing the collision to the time when the follower vehicle gets the warning. For the data collection application, we have used the following measures to evaluate both DSRC and the 5G mm-wave.

- Mean Packet Size (bytes): Total number of received bytes of each data packet
- Packet Loss Ratio (%) = $\frac{\sum_{i=1}^{n} L_i}{\sum_{i=1}^{n} S_i}$

Ratio of number of lost packets (L) for a receiver and total number of packets (S) sent to the receiver, and i is the simulation second

- Mean Delay (ms) = $\frac{\sum_{i=1}^{n} D_i}{P_i}$

  Average delay (D) per received packet for a receiver where P is the total number of packets sent during the simulation run
- Receiver Bitrate (Kbps) = $\frac{\sum_{i=1}^{n}(R_i * 8)}{N}$

  Received packet bitrate, where R is the byte received per i-th second, and N is the total simulation time
- Transmission Bitrate (Kbps) = $\frac{\sum_{i=1}^{n}(T_i * 8)}{N}$

  Transmitted packet bitrate, with T is the byte transmitted per i-th second, and N is the total simulation time

- Signal-to-interference-plus-noise ratio (SINR, db) = $\frac{M}{I+O}$

  M is the incoming signal's power, I is the interference of the other connected object, and O is the noise.

In order to calculate these communication-related parameters, we have used the flow monitor in ns-3, which monitors the packet flow between the nodes (i.e., CAVs, base stations) (22).

## ANALYSIS AND RESULT

### *Forward collision warning*

In this safety application, we have sent one forward collision warning from the suddenly stopped leading CAV to the follower CAV. In the DSRC environment, the data will be communicated directly to the follower CAV, whereas in the 5G mm-wave the data will be communicated via the mm-wave base station. We have considered, the distance between the CAVs and mm-wave base station to be 1450 ft. 5G mm-wave is very sensitive to line-of-sight. For this research, we have considered a clear line-of-sight between the base station and CAVs. In order to achieve clear line-of-sight, the base stations can be placed on top of building roofs in an urban area.

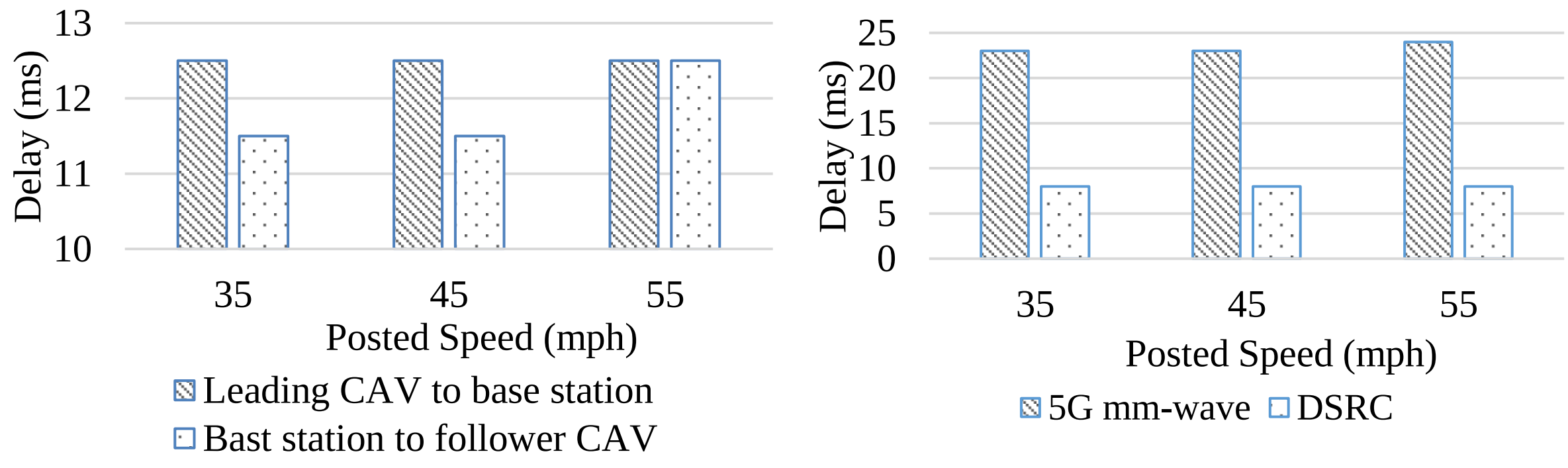

(a) Inter base station – CAV delay in 5G (b) End-to-end delay for 5G and DSRC

**FIGURE 3 Delay of DSRC and 5G-mmwave**

Figure 3 shows the: (a) inter base station – CAV delay for 5G and (b) end-to-end delay of both 5G mm-wave and DSRC for different vehicle speeds. The posted speed for Woodruff Road is 45 mph, so we tested for 35, 45 and 55 mph CAV speed. Based on Figure 3(b) for different CAV speeds, both DSRC and 5G mm-wave satisfy the forward collision application latency threshold, which is lower than 200 ms. The end-to-end delay of sending forward collision warnings by DSRC is not affected by vehicle speed. However, the end-to-end delay for the 5G mm-wave for the forward collision warning increases by 1 ms at the 55 mph speed limit compared to lower speeds, which is negligible.

*Traffic data collection*

The CAV data collection comparison between DSRC and 5G mm-wave are not conducted using the same experiments (i.e., communication data rate), as DSRC has a limited data rate compared to 5G. The analysis of both environments are discussed independently.

**DSRC-enabled CAV** In this simulation environment, we have determined the application data rate based on the standard Basic Safety Message data size and transmission rate criteria, which is 200 bytes and one-tenth of a second, respectively. Findings from our experiments as summarized in Figure 4 suggest that with more CAVs and different speeds, the mean delay does not change. The packet loss ratio is 8 to 10% for all scenarios due to the dynamic topology of the network. The transmission bitrate is almost the same for all scenarios as shown in Figure 4. The mean delay stays less than 750 ms.

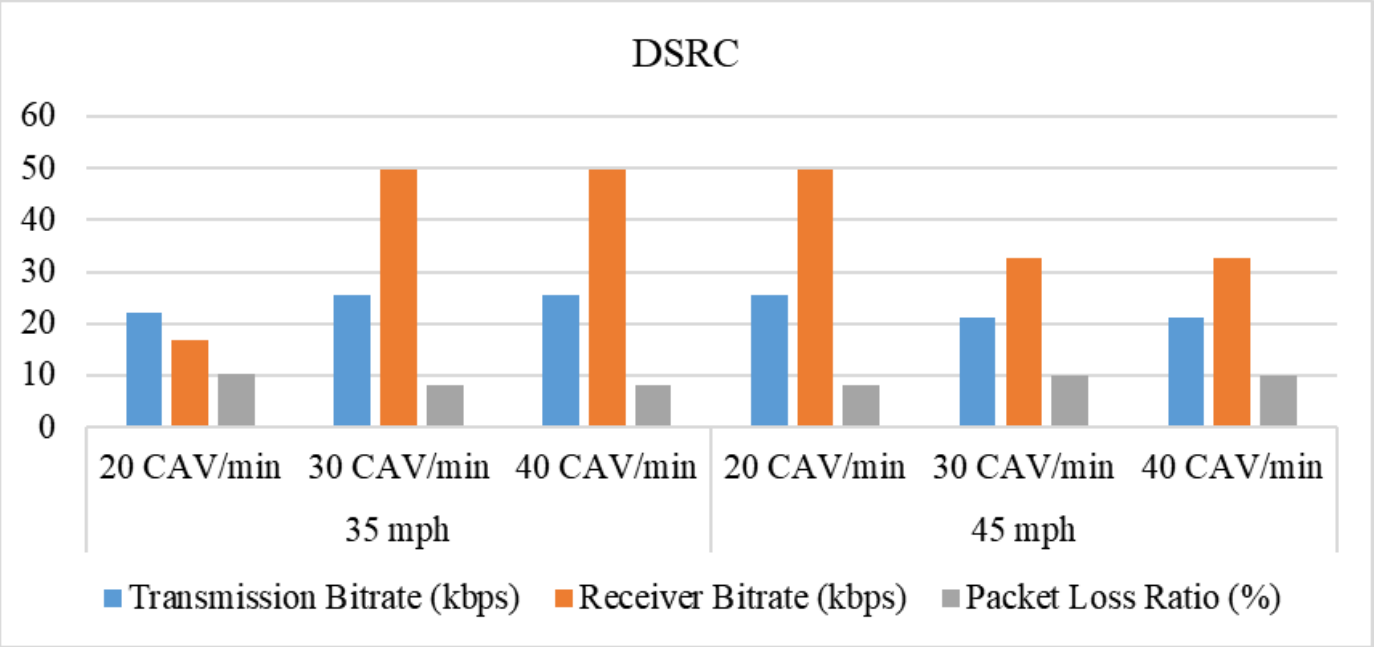

(a) DSRC bitrate and packet loss

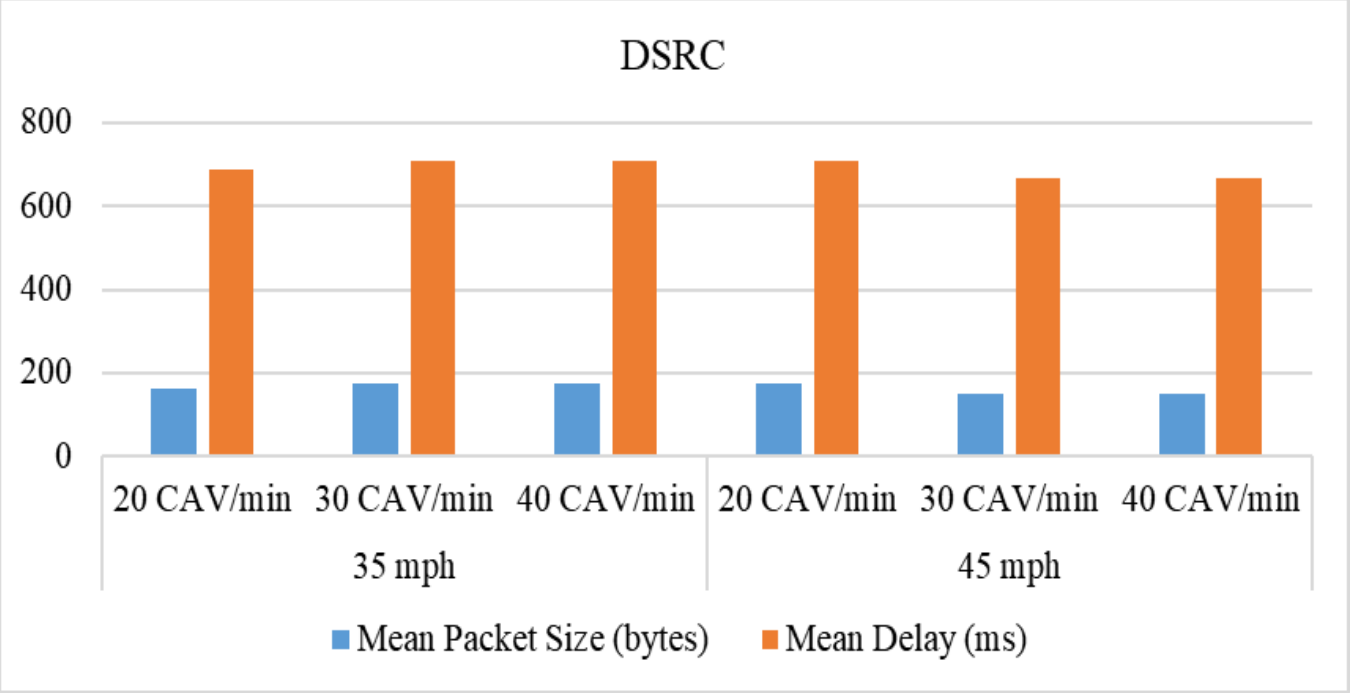

(b) DSRC average packet size and delay

**FIGURE 4 CAV data collection with DSRC**

**5G mm-wave-enabled CAV** In this simulation environment, we have used 4000 Kbps as the application data rate. Figure 5 suggests that the with the 5G mm-wave, we can achieve high reliability in the wireless communication system. The packet loss ratio is less than 1% with 20 or more CAVs/min for both posted speeds. The edge-to-edge (i.e., between the CAVs and base stations) mean delay increases with an increasing number of CAVs. This means that with multiple CAVs the end-to-end delay will not be affected for the 5G mm-wave. The receiver Bitrate is 3811 kbps for 20 CAVs/min, which satisfies the required Bitrate requirement to stream a single high definition video stream at 720p with 1280x720 resolution, which is 1,200 - 4,000 Kbps (23). Thus, 5G mm-wave can support in-vehicle infotainment systems for congested areas with multiple CAVs, which is not possible with DSRC.

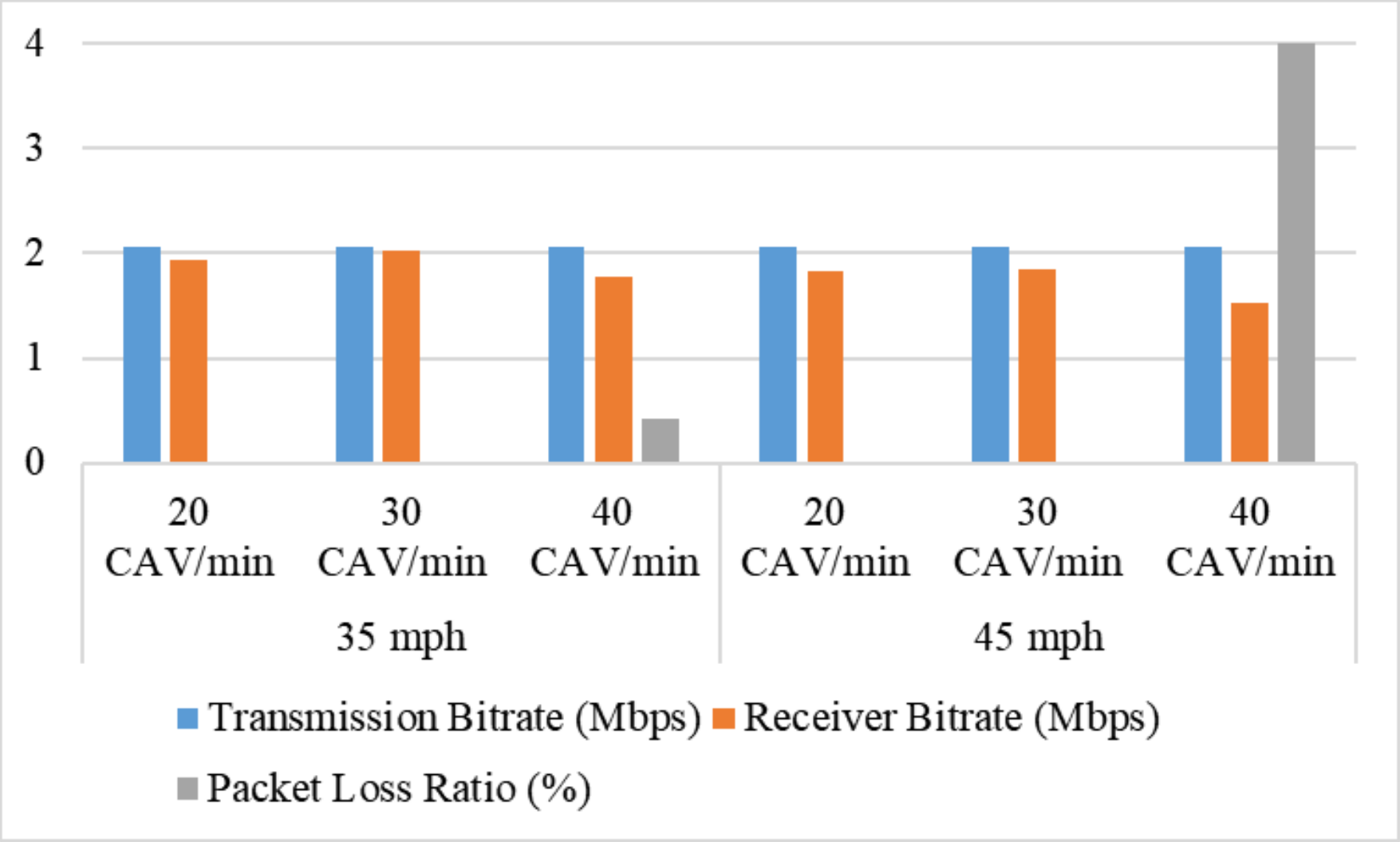

(a) 5G mm-wave bitrate and packet loss

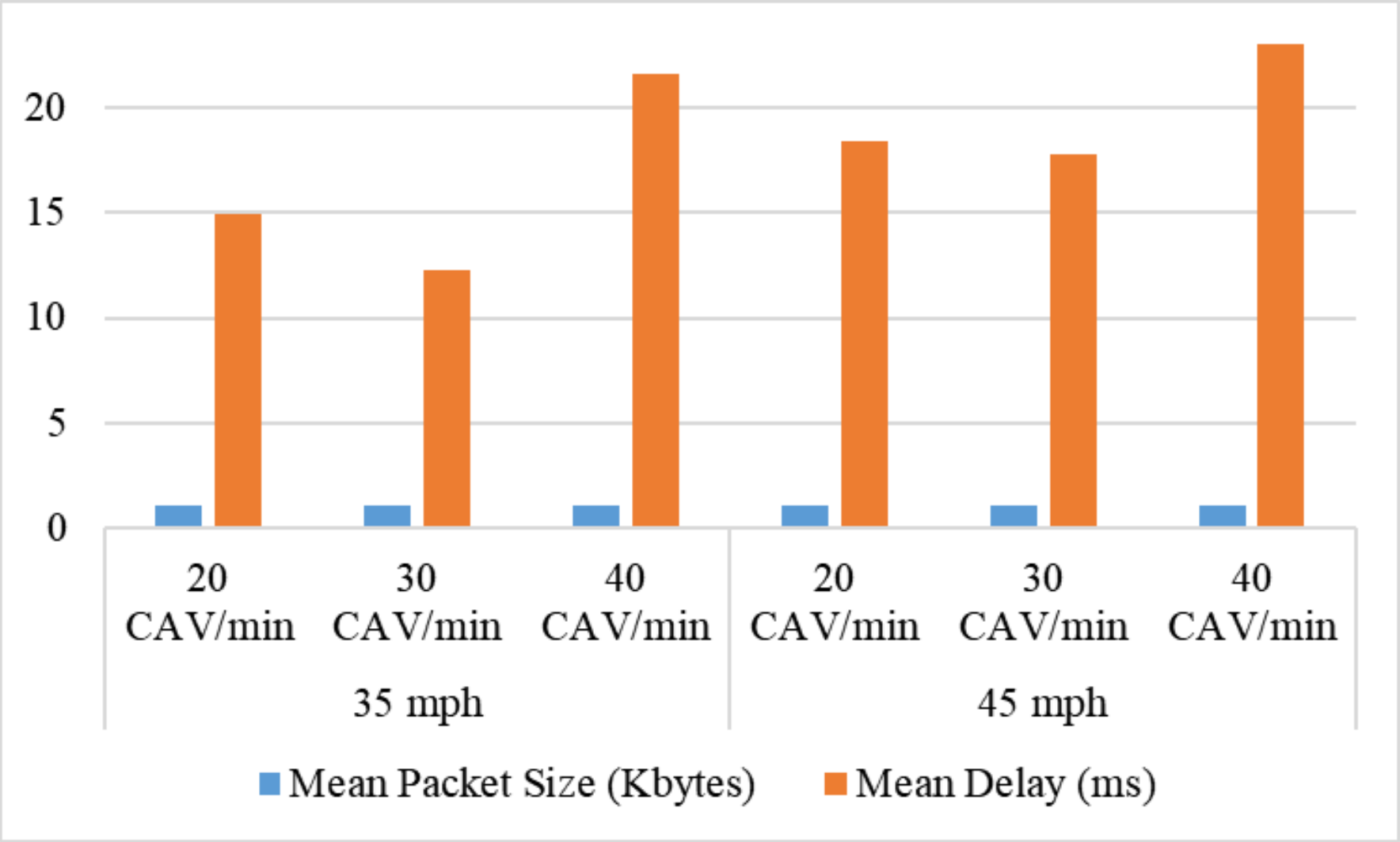

(b) 5G mm-wave average packet size and delay

**FIGURE 5 CAV data collection with 5G mm-wave**

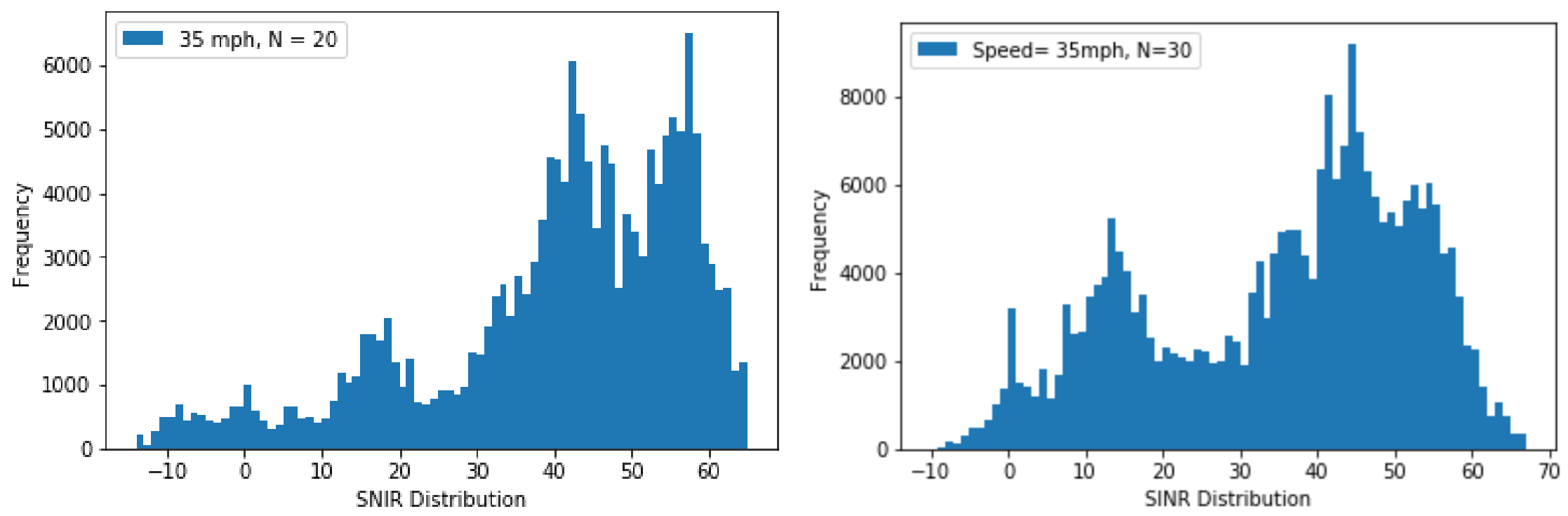

(a) 5G mm-wave SINR distribution for 35 mph

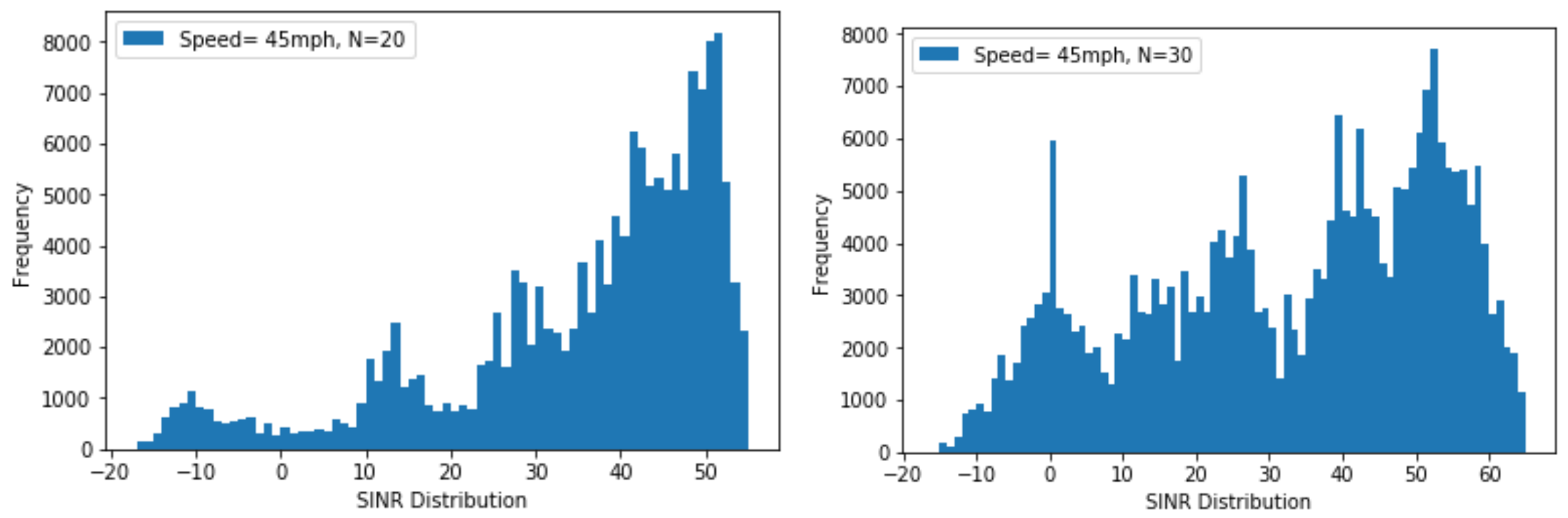

(b) 5G mm-wave SINR distribution for 45 mph

**FIGURE 6 SINR distribution of 5G mm-wave**

We have also measured the quality of 5G communication in terms of SINR. SINR is a wireless quality indicator where a higher value represents a higher quality of wireless communication with higher throughput. Figure 6 shows the SINR distribution of different scenarios. In general, the trend for different speed limits is the same. In all other cases, the most frequent SINR value observed is around 50dB. The impact of velocity is negligible in the quality of 5G. In DSRC communication, the typical threshold of SINR is 25 dB (24), whereas 5G shows a double improvement in terms of wireless communication link quality in the majority of the cases.

Based on the analysis, a summary comparison of DSRC and 5G mm-wave is presented in Table 3. As shown in Table 3, 5G mm-wave across all applications considered in this study (i.e., safety, infotainment and traffic data collection).

**Table 3: Feasibility of 5G mm-wave and DSRC**

| **Wireless communication** | **Safety application** | **Infotainment** | **Traffic data collection application** |
|---|---|---|---|
| DSRC | Feasible | Not feasible | Feasible |
| 5G mm-wave | Feasible | Feasible | Feasible |

## CONCLUSIONS

CAVs need reliable and fast communication with a high data rate to support different CAV applications, which include safety, mobility, and infotainment applications. Meeting all of these requirements has been a major challenge for existing widely deployed wireless communication services, such as 4G LTE. 5G mm-wave communication is a promising wireless communication option to support CAV applications, especially in a congested area. We have found that increasing the number of CAVs does not deteriorate the end-to-end delay for 5G mm-wave. The end-to-end delay derived from the forward collision application shows that 5G mm-wave satisfies the maximum delay threshold, for safety-critical applications, of 200 ms. As cellular service providers are showing interest in including 5G mm-wave, it will be a viable option for public agencies to include 5G mm-wave supported CAV applications.

Although mm-wave is showing promising results, their signal is susceptible to blockage and non-line-of-sight, which is why the base stations have to be closely spaced. Having closely spaced mm-wave base stations is an ideal scenario for an urban environment where CAVs will operate by communicating with other CAVs, connected pedestrians, and other external data sources. Future studies should consider multiple external data sources, including real-time data feed from traffic cameras, traffic management centers, social media, cloud database etc. Additionally, testing on the performance of a heterogeneous 5G LTE and mm-wave enabled network to provide a reliable communication system for a long coverage area, such as interstates, should be conducted.

## AUTHOR CONTRIBUTION STATEMENT

The authors confirm contribution to the paper as follows: study conception and design, Sakib Mahmud Khan, Mashrur Chowdhury; data collection, Sakib Mahmud Khan; interpretation of results, Sakib Mahmud Khan, Mashrur Chowdhury, Mhafuzul Islam; draft manuscript preparation, Sakib Mahmud Khan, Mizanur Rahman, Mhafuzul Islam, Mashrur Chowdhury. All authors reviewed the results and approved the final version of the manuscript.